\documentclass[twocolumn,prd,showpacs,showkeys,nofootinbib]{revtex4-1}
\usepackage[T1]{fontenc}
\usepackage[utf8]{inputenc}
\usepackage[english,russian]{babel}
\usepackage{float}
\usepackage{epstopdf}
\usepackage{amsmath,amssymb,pgf,graphicx,color,url,array}
\usepackage{tabularx}
\usepackage{wasysym}
\usepackage{longtable}
\newcommand{\be}{\begin{equation}}
\newcommand{\ee}{\end{equation}}
\newcommand{\todo}[1]{{}}

\usepackage[inline]{enumitem}

\def\itemrange#1{%
\addtocounter{enumi}{1}%
\edef\labelenumi{\theenumi--\noexpand\theenumi.}%
\addtocounter{enumi}{-1}%
\addtocounter{enumi}{#1}%
\item
\def\labelenumi{\theenumi.}}

\begin{document}
\vskip -0.9cm
\title{Search for Ultra-High-Energy Neutrinos with the Telescope Array Surface Detector}

\author{ R.U. Abbasi$^{1}$, M.Abe$^{2}$, T. Abu-Zayyad$^{1}$, M. Allen$^{1}$, E. Barcikowski$^{1}$, J.W. Belz$^{1}$, D.R. Bergman$^{1}$, S.A. Blake$^{1}$, R. Cady$^{1}$, B.G. Cheon$^{4}$
% ...
\\~\\
{\footnotesize\it
$^{1}$ High Energy Astrophysics Institute and Department of Physics and Astronomy, University of Utah, Salt Lake City, Utah, USA \\
$^{2}$ The Graduate School of Science and Engineering, Saitama University, Saitama, Saitama, Japan \\
$^{3}$ Graduate School of Science and Engineering, Tokyo Institute of Technology, Meguro, Tokyo, Japan \\
$^{4}$ Department of Physics and The Research Institute of Natural Science, Hanyang University, Seongdong-gu, Seoul, Korea \\
$^{5}$ Department of Physics, Tokyo University of Science, Noda, Chiba, Japan \\
$^{6}$ Department of Physics, Kindai University, Higashi Osaka, Osaka, Japan \\
$^{7}$ Service de Physique ThÃ©orique, UniversitÃ© Libre de Bruxelles, Brussels, Belgium \\
$^{8}$ The Hakubi Center for Advanced Research and Graduate School of Science, Kyoto University, Kitashirakawa-Oiwakecho, Sakyo-ku, Kyoto, Japan \\
$^{9}$ Graduate School of Science, Osaka City University, Osaka, Osaka, Japan \\
$^{10}$ Institute for Cosmic Ray Research, University of Tokyo, Kashiwa, Chiba, Japan \\
$^{11}$ Kavli Institute for the Physics and Mathematics of the Universe (WPI), Todai Institutes for Advanced Study, University of Tokyo, Kashiwa, Chiba, Japan \\
$^{12}$ Information Engineering Graduate School of Science and Technology, Shinshu University, Nagano, Nagano, Japan \\
$^{13}$ Faculty of Engineering, Kanagawa University, Yokohama, Kanagawa, Japan \\
$^{14}$ Interdisciplinary Graduate School of Medicine and Engineering, University of Yamanashi, Kofu, Yamanashi, Japan \\
$^{15}$ Astrophysical Big Bang Laboratory, RIKEN, Wako, Saitama, Japan \\
$^{16}$ Department of Physics, Sungkyunkwan University, Jang-an-gu, Suwon, Korea \\
$^{17}$ Department of Physics, Tokyo City University, Setagaya-ku, Tokyo, Japan \\
$^{18}$ Institute for Nuclear Research of the Russian Academy of Sciences, Moscow, Russia \\
$^{19}$ Advanced Research Institute for Science and Engineering, Waseda University, Shinjuku-ku, Tokyo, Japan \\
$^{20}$ Department of Physics, Chiba University, Chiba, Chiba, Japan \\
$^{21}$ Department of Physics, School of Natural Sciences, Ulsan National Institute of Science and Technology, UNIST-gil, Ulsan, Korea \\
$^{22}$ Department of Physics, Yonsei University, Seodaemun-gu, Seoul, Korea \\
$^{23}$ Academic Assembly School of Science and Technology Institute of Engineering, Shinshu University, Nagano, Nagano, Japan \\
$^{24}$ Faculty of Science, Kochi University, Kochi, Kochi, Japan \\
$^{25}$ Nambu Yoichiro Institute of Theoretical and Experimental Physics, Osaka City University, Osaka, Osaka, Japan \\
$^{26}$ Department of Physical Sciences, Ritsumeikan University, Kusatsu, Shiga, Japan \\
$^{27}$ Sternberg Astronomical Institute, Moscow M.V. Lomonosov State University, Moscow, Russia \\
$^{28}$ Department of Physics and Astronomy, Rutgers University - The State University of New Jersey, Piscataway, New Jersey, USA \\
$^{29}$ Earthquake Research Institute, University of Tokyo, Bunkyo-ku, Tokyo, Japan \\
$^{30}$ Department of Engineering Science, Faculty of Engineering, Osaka Electro-Communication University, Neyagawa-shi, Osaka, Japan \\
$^{31}$ Graduate School of Information Sciences, Hiroshima City University, Hiroshima, Hiroshima, Japan \\
$^{32}$ Institute of Particle and Nuclear Studies, KEK, Tsukuba, Ibaraki, Japan \\
$^{33}$ National Institute of Radiological Science, Chiba, Chiba, Japan \\
$^{34}$ CEICO, Institute of Physics, Czech Academy of Sciences, Prague, Czech Republic \\
$^{35}$ Department of Physics and Institute for the Early Universe, Ewha Womans University, Seodaaemun-gu, Seoul, Korea \\
$^{36}$ Department of Physics, Ehime University, Matsuyama, Ehime, Japan\\
$^{37}$ Faculty of Physics, M.V. Lomonosov Moscow State University, Moscow, Russia
}}\let\thefootnote\relax\footnote{$^{*}$ Deceased}
\let\thefootnote\relax\footnote{$^{**}$ Corresponding author, zhezher.yana@physics.msu.ru}
\let\thefootnote\relax\footnote{$^{***}$ Now at INFN, sezione di Torino, Turin, Italy}

\begin{abstract}
We present an upper limit on the flux of ultra-high-energy down-going neutrinos for $E > 10^{18}\ \mbox{eV}$ derived with the nine years of data collected by the Telescope Array surface detector (05-11-2008 -- 05-10-2017). The method is based on the multivariate analysis technique, so-called Boosted Decision Trees (BDT). Proton-neutrino classifier is built upon 16 observables related to both the properties of the shower front and the lateral distribution function.
\end{abstract}
\maketitle

\section{Introduction}
\label{sec:intro}

The Telescope Array (TA) experiment is the largest ultra-high-energy (UHE)
cosmic-ray experiment in the Northern Hemisphere, located near Delta,
Utah, USA~\cite{Tokuno}. TA is designed to register the extensive air showers (EAS), cascades of secondary particles produced in the interactions of cosmic rays with energies greater than $10^{18}\ \mbox{eV}$ with the Earth's atmosphere. In the Telescope Array, air showers are registered in two ways: particle density and the shower timing on the ground are measured with the surface detector (SD) array~\cite{TASD}, while the fluorescence light from gas molecules in the atmosphere excited and ionized by the passage of EAS particles is detected with 38 fluorescence telescopes grouped into three fluorescence detector stations -- Middle Drum, Black Rock Messa and Long Ridge~\cite{Tokuno2}. The simultaneous use of both surface detectors and fluorescence telescopes is known as the hybrid technique.

The surface detector facility is an array of 507 plastic scintillator stations arranged on a square grid
with 1.2 km spacing covering an approximate area of 700 ${\mbox{km}}^2$. Each detector is composed of two layers of 1.2 cm thick extruded scintillator of the $3\ {\mbox{m}}^2$ effective area.

Ultra-high-energy cosmic rays pose a number of long-standing questions in astrophysics: neither their origin, nor the acceleration mechanisms responsible for the observed UHECR flux have been identified so far. Recent observation of dipole anisotropy by the Pierre Auger Observatory~\cite{Aab:2017tyv} indicates that highest-energy cosmic rays are born in the extragalactic sources. Because of their properties, cosmic-ray neutrinos are thought to be one of the efficient tools to study UHECR models~\cite{Aloisio:2015ega,vanVliet:2017obm,Heinze:2015hhp,Giacinti:2015pya}. Neutrinos and photons are produced in hadron-hadron and photon-hadron interactions through the production and decay of pions and kaons.  In contrast to photons, a negligible fraction of neutrinos is absorbed during their propagation to Earth due to the small cross-sections of its interactions with other particles. Neutrinos are not deflected by magnetic field which means they point directly to their sources and may give a hint on the type of UHECR sources and on the mechanisms responsible for the acceleration of a parent particle~\cite{Waxman:1998yy,Waxman:1997ti,Engel:2001hd,Murase:2010gj,Aartsen:2016ngq}. Moreover, if the primary particles are not purely protons, the neutrino flux will be suppressed considerably, so they may also be a probe of a mass composition of the UHECR~\cite{Aloisio:2017waw,vanVliet:2019nse}.

UHE neutrinos may be born in the Universe in three different types of sources and processes:

\begin{enumerate}
\item \textit{Astrophysical} neutrinos are born in the hadron interactions of the UHECR with radiation or matter near its astrophysical sources. Most promising among the sources are active galactic nuclei (AGN) and its sub-class, blazars, where neutrinos are born in the proton interactions with the AGN photon fields~\cite{Stecker:1991vm}; galaxy clusters~\cite{Berezinsky:1996wx} and starburst galaxies~\cite{Thompson:2006np}, whose intergalactic and interstellar medium serves as a target for UHECR; supernovae~\cite{Budnik:2007yh} and hypernovae~\cite{Wang:2007ya}, suggested for a long time as the cosmic-ray accelerators; and gamma-ray bursts (GRBs)~\cite{Waxman:1997ti}.

\item Primary particles and nuclei may interact with the cosmic microwave background (CMB) and
the extragalactic background light (EBL) during their propagation in the Universe. Primary protons undergo photo-pion production ($p + \gamma \rightarrow p/n + \pi^0/\pi^+$) process, while the relevant process for nuclei is the photodisintegration ($\left(A,Z \right) + \gamma \rightarrow \left(A',Z' \right) + \left(Z - Z' \right)p + \left(A - A' +  Z - Z'\right)n$). The photo-hadronic interactions at the highest energies should result in the flux of the so-called \textit{cosmogenic} neutrinos~\cite{Berezinsky:1970xj,Stecker:1973sy,Hill:1983xs}.

\item Lastly, UHE neutrinos may be born in top-down models in decays of massive objects, such as heavy dark matter particles and topological defects~\cite{Berezinsky:1991aa,Bhattacharjee:1991zm,AlvarezMuniz:2000es,Gelmini:1999ds,Kusenko:2000fk} in the processes such as $D \rightarrow \nu + all$, or a possible rare decay $D \rightarrow 3 \nu$.
\end{enumerate}

Multiple methods have been used to search for UHE neutrinos~\cite{Baret:2011zz,Anchordoqui:2009nf}. Neutrinos may interact with the Earth's atmosphere through charged ($\nu + N \rightarrow \mbox{lepton} + X $) and neutral ($\nu + N \rightarrow \nu + X$) current interactons (CC and NC from now on). Electrons and tau-leptons produced in NC interactions as well as final-state products $X$ of hadronic interactions generate extensive air showers, which, in turn, may be detected with ground arrays, fluorescence telescopes and radio antennas. Such neutrino events are usually called \textit{down-going}.

However, neutrino-air interaction cross-sections are much smaller than the ones for protons or other nuclei~\cite{Kusenko:2001gj} thus the probability to invoke an EAS is also suppressed. To overcome this obstacle, it was suggested in~\cite{Beresinsky:1969qj,Berezinsky:1975zz}, that air showers initiated by the UHE neutrinos may be observed at large zenith angles, which increases the slant depth neutrinos travel in the atmosphere. The probability of neutrino-air interaction is constant at any point along their trajectories, and this allows to distinguish neutrino-induced showers as the ones which develop deep in the atmosphere, unlike the protons and nuclei showers which are initiated in the upper layers.

Another possibility for neutrino to invoke an EAS is the CC interactions with the minerals in the Earth's crust -- so-called \textit{Earth-skimming} events. They usually occur close to the exit point below the surface of the Earth and the EAS develops upwards in the atmosphere. Whilst muon and electron neutrinos produce
well-contained tracks and cascades after one CC-interaction, ${\nu}_{\tau}$ generates a $\tau$-lepton, and a characteristic interaction length of such process is about $20\ \mbox{km}$. The decay length of a $\tau$-lepton is then sufficient enough for it to escape the Earth's crust and generate an EAS~\cite{Bottai:2002nn,Feng:2001ue}.

Thirdly, one can observe the radio-emission from neutrino passing through dense matter, such as ice or lunar regolith, caused by the Askaryan effect~\cite{Askaryan}. In the Askaryan effect, the coherent radiowave Cherenkov radiation is initiated by the particles moving through the medium with speeds close to the speed of light.

Previously, cosmic-ray neutrino searches have been performed by various experiments. At the ultra-high energies, neutrino studies were performed by the Fly's Eye~\cite{Baltrusaitis:1985mt}, the HiRes~\cite{Abbasi:2008hr}, the Westerbork Synthesis Radio Telescope~\cite{Scholten:2009ad}, the RESUN project~\cite{Jaeger:2009zb}, the LUNASKA~\cite{James:2009sf,Bray:2015lda} and RICE~\cite{Kravchenko:2011im} experiments, the Pierre Auger Observatory~\cite{Aab:2015kma,Aab:2019auo}, the ARIANNA array~\cite{Barwick:2014pca} and the ANITA~\cite{Gorham:2008yk}, ANITAII~\cite{Gorham:2010} and ANITAIII balloon payloads~\cite{Allison:2018cxu}. No neutrinos with energies more than $10^{16}\ \mbox{eV}$ have been detected so far and only the upper limits on the differential flux were derived.

At the TeV to PeV energies, results were obtained with the Askaryan Radio Array (ARA)~\cite{Allison:2015eky,Allison:2019xtn}, the ANTARES telescope~\cite{Albert:2017oba} and the IceCube observatory~\cite{Albert:2018vxw}. An astrophysical neutrino flux above the atmospheric neutrino background was discovered by IceCube~\cite{Aartsen:2013jdh} in the three years of its data. The latest measured astrophysical neutrino flux based on the IceCube six-year data is comprised of 82 events~\cite{Aartsen:2017mau}, including two neutrinos with energies above 1 PeV, and a 2 PeV event~\cite{Aartsen:2014gkd}.

The present \textit{Paper} is dedicated to the ultra-high-energy down-going neutrino search using solely the data from the Telescope Array Surface Detector array~\cite{TASD}. The data from the nine years of the Telescope Array surface detector operation from May 11, 2008 to May 10, 2017 are used in the present analysis. Fluorescence detectors operate only on clear moonless nights, which results in approximately 10 \%\ duty cycle, while the SD array duty cycle reaches 95 \%\ which make the use of its data much more beneficial in terms of acquired statistics.

While there are no observables, which can be obtained from the SD data and are as sensitive to the composition of primary particles, as the depth of the shower maximum, $X_{\mbox{max}}$, an alternative approach was proposed. The data analysis employs the multivariate analysis techniques to obtain the proton-neutrino Boosted Decision Trees (BDTs) classifier~\cite{Breiman,Schapire} built upon 16 observables. For each event, a classifier returns a single value $\xi$, which is then available to one-dimensional anaylsis and allows one to distinguish between different hypotheses.

The {\it Paper} is organized as follows: in the Section~\ref{sec:data} data and Monte-Carlo sets are described. In the Section~\ref{sec:method} multivariate analysis method is addressed together with its implementation to the neutrino search. Finally, results and discussions are provided in Section~\ref{sec:results}.

\section{Data set and simulations}
\label{sec:data}

\subsection{Data set}
\label{subsec:sddata}

The data from the nine years of the Telescope Array surface detector operation from May 11, 2008 to May 10, 2017 are used in the present {\it Paper}. Each event is comprised of the time-dependent signals (waveforms) from both upper and lower layers of each triggered detector. The waveforms are recorded by the 12-bit flash analog-to-digital converters (FADC) with the 50 MHz sampling rate and are converted to Minimum Ionizing Particles (MIP)~\cite{TASD} at the calibration stage. The station is marked as triggered if the signal exceeds $0.3\ \mbox{MIP}$.

\subsection{Event reconstruction and cuts}
\label{subsec:reconstruction}

Each surface detector array event is comprised of a set of observables $t_i \left(r_i \right)$ and $S_i \left(r_i \right)$, where $t_i \left(r_i \right)$ is the time of arrival of an EAS to the $i$-th detector of the event and $S_i \left(r_i \right)$ is the signal from the station. These observable quantities are used to perform a joint fit of the geometry and lateral distribution function (LDF). The shower front is approximated with empirical functions proposed by Linsley~\cite{Linsley} and later modified in the AGASA experiment~\cite{Teshima}. The $S_{800}$ parameter, the signal value at $800\ \mbox{m}$ from the shower core used as an energy estimator, is then obtained using pulse heights in the counters together with the event geometry information~\cite{Takeda}.

For comparison, a number of Monte-Carlo events are generated with the CORSIKA package~\cite{Heck} and GEANT4-based detector simulator~\cite{Agostinelli}. MC events are also reconstructed with the same code and procedure for the data.

The Linsley front curvature parameter is determined through the following fit to the shower front using the LDF with 7 free parameters: $x_{\mbox{core}}$, $y_{\mbox{core}}$, $\theta$, $\phi$, $S_{800}$, $t_0$, $a$~\cite{TAgammalim}:

\begin{equation}
t \left(r \right) = t_0 + t_{\mbox{plane}} + a \times \left(1+r/R_L\right)^{1.5} LDF\left(r \right)^{-0.5},
\end{equation}
\begin{equation}
S \left(r \right) = S_{800} \times LDF\left(r \right),
\end{equation}
\begin{equation}
f\left(r \right) = \left(\frac{r}{R_m}\right)^{-1.2}\left(1+\frac{r}{R_m}\right)^{-(\eta-1.2)}\left(1+\frac{r^2}{R_1^2}\right)^{-0.6},
\end{equation}
\begin{equation*}
t_{\mbox{plane}}^i = \frac{1}{c} \vec n \left(\vec R_i - \vec R_{\mbox{core}}\right),
\end{equation*}
\begin{equation*}
R_m = 90.0~\mbox{m},~R_1 = 1000~\mbox{m},~R_L = 30~\mbox{m},
\end{equation*}
\begin{equation*}
\eta = 3.97 - 1.79 \left(\sec\left(\theta\right) -1\right),
\end{equation*}
\begin{equation*}
r = \sqrt{ (\vec R_i - \vec R_{core})^2 - (\vec n (\vec R_i - \vec R_{core}))^2},
\end{equation*}

\noindent where $\vec x_{\mbox{core}}$ and $\vec y_{\mbox{core}}$ are the locations of the shower core, $\vec x_i$ and $\vec y_i$ are the locations of each station of an event, obtained from the pre-defined coordinate system of the array centered at the Central Laser Facility (CLF)~\cite{Takahashi:2011zzd}, $t_{plane}$ is the arrival timing of the shower plane at the distance $r$, $\vec n$ -- unit vector towards the direction of arrival of a primary particle, $c$ is a speed of light and $a$ is the Linsley front curvature parameter.

The following cuts are applied:

\begin{enumerate}
\item the event includes 5 or more triggered stations;
\item zenith angle $\theta \in [45^{\circ};90^{\circ}]$;
\item reconstructed core position inside the array with
  the distance of at least $1200\ \mbox{m}$ from the edge of the array;
\item Effective ${\chi}^2 / \mbox{d.o.f.}$ doesn't exceed 5 for the joint geometry and LDF fit.
\end{enumerate}

No energy cut is applied, and after the cuts, the SD data set contains 197250 events.

\subsection{Simulations}
\label{subsec:simulations}

For the Monte-Carlo simulations, the CORSIKA software package~\cite{Heck}
is used along with the QGSJETII-03 model for high-energy hadronic
interactions~\cite{Ostapchenko}, FLUKA~\cite{FLUKA,FLUKA2} for low energy hadronic
interaction and EGS4~\cite{EGS4} for electromagnetic processes. Interactions of primary neutrinos are handled with HERWIG~\cite{Corcella:2000bw} integrated with CORSIKA. While a newer version of the QGSJET model, QGSJETII-04~\cite{QGSJETII:04}, is also widely incorporated in the related analyses, the previous version is found to be adequate to describe the data~\cite{Abbasi:2018wlq}; and an upper limit on the neutrino flux is expected to be conservative in respect to the hadronic interaction models as discussed in Section~\ref{sec:results}.

The neutrino MC set is comprised of 3000 CORSIKA showers generated in the energy range $3 \times 10^{17} - 3 \times 10^{20}\ \mbox{eV}$. Due to the low neutrino-interaction cross-section a primary neutrino thrown with CORSIKA will most likely travel through the Earth's atmosphere without interaction. To overcome this obstacle the height of the primary interaction is simulated with a separate routine and fixed manually within CORSIKA.

The procedure consists of the following steps: at first, a uniform neutrino flavour flux ${\nu}_{e}:\bar{{\nu}}_{e}:{\nu}_{\mu}:\bar{{\nu}}_{\mu}:{\nu}_{\tau}:\bar{{\nu}}_{\tau} = 1 : 1 : 1 : 1 : 1 : 1$ is assumed and a randomly chosen flavour is assigned for the event. Then, zenith angle $\theta \in [0;90^{\circ}]$ is thrown assuming an isotropic flux. For a given zenith angle, the slant depth of the atmosphere $T_{atm}$ is calculated.

Then, we continue with an event if the interaction happens in the atmosphere. The probability of the latter is given by $T_{\mbox{atm}} / T_{\mbox{int}}$, where $T_{\mbox{int}} = \left( M / N_A \right) / \sigma_{CC+NC}$ is the interaction slant depth in low interaction probability approximation $T_{atm} / T_{int} \ll 1$, where $M$ is the average molar mass of air, $N_A$ is the Avogadro number. Neutrino-nucleus interaction cross-sections $\sigma_{CC+NC}$ are taken from~\cite{CooperSarkar:2007cv}.

Finally, the interaction slant depth is thrown in the range $[0;T_{atm}]$ and the derived value is fixed within CORSIKA input file. Simulated neutrino showers were used to throw 80 millions of events on the detector by randomly distributing the core location and azimuthal angle.

The highly-inclined proton MC set consists of 2400 proton showers with $45^{\circ} < \theta < 90^{\circ}$ in the energy range $3 \times 10^{17} - 3 \times 10^{20}\ \mbox{eV}$ used to throw 100 millions of events of each type on the SD array.

For both neutrino and highly-inclined proton MC sets, the usual CORSIKA treatment of atmosphere as a flat disk becomes inefficient at high zenith angles, and one needs to take the curvature of the atmosphere into account. This is done within CORSIKA with a special CURVED atmosphere option~\cite{Heck}.

CPU-time-saving \textit{thinning} procedure is utilized for both of the MC sets~\cite{Hillas}. Instead of following each of millions of particles born in an EAS, the method follows in detail only those of the particles which have an energy above a certain threshold, which is specified by a fraction of the primary energy $\epsilon_{th}$. Below this limit, only one particle out of the secondaries produced in a certain interaction is randomly selected and assigned with a weight to ensure energy conservation. In the present work, the thinning level of
$\epsilon_{th}=10^{-6}$ is used.

Statistical properties of a shower are restored with a \textit{dethinning} procedure~\cite{Stokes}. The detector response is simulated by the GEANT4 package~\cite{Agostinelli}. Real-time array status and detector calibration information for nine years of observations are used for each simulated event~\cite{TAdataMC}.


\subsection{Composition-sensitive observables}
\label{subsec:observables}

During the reconstruction procedure, the following composition-sensitive observables are obtained for each data and MC event:

\begin{enumerate}[series=MyList]
\item Linsley front curvature parameter, as described in section \ref{subsec:reconstruction}.
\itemrange{1} Area-over-peak (AoP) of the signal at 1200 m and AoP slope parameter~\cite{Abraham}:
\end{enumerate}

\begin{changemargin}{0.75 cm}{0 cm}
$AoP\left(r \right)$ is fitted with a linear fit:

\begin{equation*}
AoP\left(r \right) = \alpha - \beta \left( r/r_0 - 1.0 \right),
\end{equation*}

where $r_0=1200$\,m, $\alpha$ is $AoP\left(r \right)$ value at 1200 m
and $\beta$ is its slope parameter.

\end{changemargin}

\begin{enumerate}[resume=MyList]
\item Number of detectors hit.
\item Number of detectors excluded from the fit of the shower front by the reconstruction procedure~\cite{AbuZayyad3}.
\item Effectuve $\chi^2/ \mbox{d.o.f.}$
\itemrange{1} $S_b$ parameter for $b=3$ and $b=4.5$~\cite{Ros}. The definition of
the parameter is the following:

\begin{equation*}
S_b = \sum_{i=1}^{N} \biggl[ S_i \times {\left( \frac{r_i}{r_0} \right)}^b \biggr],
\end{equation*}

\noindent where $S_i$ is the signal of {\it i}-th detector, $r_i$ is the
distance from the shower core to this station in meters and $r_0  =
1200\ \mbox{m}$ -- reference distance. The value $b = 3$ and $b = 4.5$
are used as they provide the best separation.

\item The sum of the signals of all the detectors of the event.
\item Asymmetry of the signal at the upper and lower layers of detectors.
\item Total number of peaks within all FADC (flash analog-to-digital converter) traces. Number of peaks is averaged over both upper and lower layers of all stations participating in the event. FADC peak is defined as a time bin with a signal above 0.2 Vertical equivalent muons (VEM) higher than signal in the 3 preceding and 3 following time bins.
\item Number of peaks for the detector with the largest signal.
\item Number of peaks present in the upper layer and not in the lower.
\item Number of peaks present in the lower layer and not in the upper.
\end{enumerate}

Together with the zenith angle and $S_{800}$, the aforementioned variables comprise a set of 16 composition-sensitive observales used to build a multivariate classifier. They are discussed in more detail in~\cite{Abbasi:2018wlq}.




\section{Method}
\label{sec:method}

\subsection{Boosted Decision Trees}
\label{subsec:BDT}

The Boosted Decision Tree (BDT) concept belongs to a number of so-called multivariate analysis techniques which allow to efficiently treat the data described by vector-type variables. BDTs are used in a variety of problems, where it is necessary to define, whether a particular event belongs to the signal or to the background. Boosted Decision Trees were previously succesfully used in a number of problems related to cosmic-ray data analysys~\cite{Abbasi:2018wlq,Abbasi:2018ywn,Abbasi:2019mwi}.

In general, a single classifier, a \textit{tree} is built as follows:

\begin{enumerate}
  \item For each variable a splitting value with best separation is found. This value divides the full range of the values of the variable into two ranges, which are called \textit{branches}. It will be mostly signal in one branch, and mostly background in another;
  \item The algorithm is repeated recursively on each branch. It may use a new variable or reuse the same one;
  \item The decision tree will iterate until the stopping criterion is reached (for example, number of events in a branch). The terminal node is called a \textit{leaf}.
\end{enumerate}

A single classifier may not be efficient enough to provide a good separation between signal and background events, in this case it is called \textit{weak}. Instead, one may build a number of weak classifiers -- a \textit{forest} to create a strong one. This is the idea which is the basis of the concept of \textit{boosting}. The adaptive boosting (AdaBoost) algorithm is employed in the present work~\cite{Schapire,Freund} with the number of trees $N_{\rm Trees}=1000$. In AdaBoost, a weak classifier is run multiple times on the training data, and each event is weighted by how incorrectly it was classified. An improved tree with reweighted events may now be built, and as a result, averaging over all trees allows the creation of a better classifier.

The method finally results in a single value $\xi$ for each event. $\xi$ resides in the range $\xi \in [-1;1]$,
where $\xi = 1$ corresponds to a pure signal event , $\xi = -1$ -- pure background event.

In this \textit{Paper}, BDT method available as a part of the ROOT~\cite{root} Toolkit for Multivariate Data Analysis (TMVA) package~\cite{Hocker} is used. The neutrino and highly-inclined proton Monte-Carlo sets are split into three and two parts with equal
statistics, correspondingly. The first part of both sets is used to build and train the BDT classifier based on 16 variables listed in~\ref{subsec:observables}. Highly-inclined-proton-induced MC showers are used as a background and neutrino-induced ones as signal events. The second part is used for the cut optimization. The third part of the neutrino MC set is used for the exposure calculation. The classifier is applied to the data set as well to the second and third parts of the Monte-Carlo sets.

The zenith angle distribution for MC events plotted against the $\xi$ parameter is shown in Figure~\ref{xiscatter_cut}, where neutrino MC is shown with green pluses, while highly-inclined proton MC is shown with red crosses. The $\left( \xi,\ \theta \right)$ points are used in the cut optimization.

\subsection{Cut optimization}
\label{subsec:FC}

The $\xi$ parameter derived in the Section~\ref{subsec:BDT} is then used for the neutrino candidate search. For this purpose, one needs to derive a cut on $\xi$ which depends on the zenith angle. The cut is derived by using the highly-inclined proton MC set as the null-hypothesis, a set without any neutrino events.

In this case, a neutrino candidate should have a $\xi$ parameter larger, than $\xi_{cut} \left( \theta \right)$, where the latter is defined with a quadratic polynomial function:

\begin{equation*}
\xi_{cut} \left( \theta \right) = \xi_0 + \xi_1 * \theta + \xi_2 * {\theta}^2.
\end{equation*}

$\xi_{cut} \left( \theta \right)$ is derived with the use of highly-inclined proton and neutrino MC sets by minimizing the merit function, which corresponds to the mean expected value of the upper limit on neutrino flux. The merit function is calculated in the following way: some initial values of $\xi_0$, $\xi_1$ and $\xi_2$ are assumed as a starting point for the optimization. Applying the ``starting'' cut $\xi_{cut}^0$ to the highly-inclined proton and neutrino MC sets one obtains the number $n_p$ and $n_{\nu}$ of events, correspondingly, which pass the applied cut.

$n_{\nu}$ is proportional to the neutrino exposure of the SD. The number of neutrino candidates is derived from $n_p$. Among the events from the ``null hypothesis'' set which pass the cut, there are both proton events which pass the cut due to the incomplete separation between proton and neutrino events by the $\xi$ variable as well as actual neutrino events. Thus, we define the number of neutrino candidates as the random Poisson variable with the number of observed events being the $n_p$, normalized to the statistics of the experiment.

Since $n_p$ is always a small value, we apply the Feldman-Cousins method~\cite{FC} to estimate the number of neutrino candidates as the 90\% C.L. upper limit on the mean value $\langle n_p \rangle$ of the Poisson variable with the number of observed events $n_p$ and the expected number of background events $b=0$.

Finally, the merit function is estimated as:

\begin{equation}
f_{merit} \left(\xi_0,\xi_1,\xi_2,\theta \right) = \frac{\langle n_p \rangle_{90 \%\ \mbox{C.L.}}}{n_{\nu}},
\end{equation}

\noindent and it is minimized with the use of simplex algorithm~\cite{Murty}.

\begin{figure}
\includegraphics[width=0.95\columnwidth]{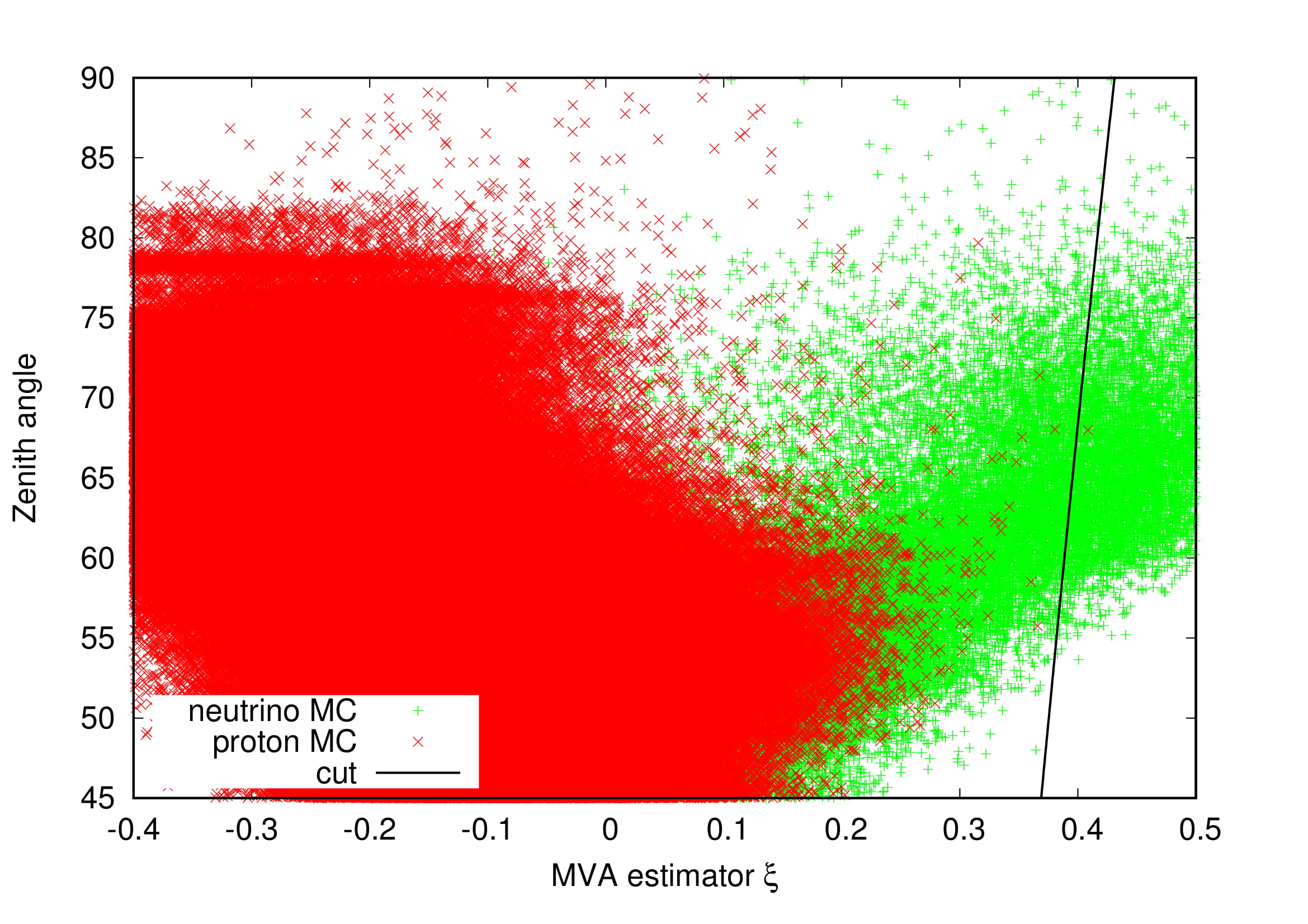}
\caption{The reconstructed zenith angle distribution for Monte-Carlo events plotted against the BDT $\xi$ estimator with the obtained $\xi_{cut}$ function ploted with black line. Neutrino MC is shown with green pluses, while highly-inclined proton MC is shown with red crosses.}
\label{xiscatter_cut}
\end{figure}

The obtained optimized $\xi_{cut} = 0.3161 + 0.0612 \times \theta + 0.0076 \times {\theta}^2$ is shown in Figure~\ref{xiscatter_cut}, where zenith angle is assumed to be in radians. 

\subsection{Exposure calculation}
\label{subsec:exposure}

The geometrical exposure for the 9 year SD observational period for the $0^{\circ} < \theta < 90^{\circ}$ is $A_{geom}^{MC} = 55500\ {\mbox{km}}^2\ \mbox{sr}\ \mbox{yr}$.

The exposure is then obtained as the ratio of number of MC neutrino events which have passed all the quality cuts and the $\xi$ cut to the number of neutrinos thrown in the atmosphere, multiplied by the geometrical exposure and the number of neutrino flavors:

\begin{equation*}
A^{\nu}_{eff} = A_{geom}^{MC} \times \frac{N_{pass}}{N_{thrown}} \times N_{flavor}.
\end{equation*}

It corresponds to the number of neutrino MC events which pass all the cuts $N_{pass} = 8460$ (21.3 \% of the number of neutrino set events which have passed the cuts). The number of events thrown in the atmosphere is $N_{thrown} = 2.81\times10^{11}$, which finally leads to the effective exposure for the down-going neutrino:

\begin{equation*}
A^{\nu}_{eff} = 4.2\times 10^{-3}\ {\mbox{km}}^2\ \mbox{sr}\ \mbox{yr}.
\end{equation*}

\section{Results and discussion}
\label{sec:results}

\begin{figure}
\includegraphics[width=0.95\columnwidth]{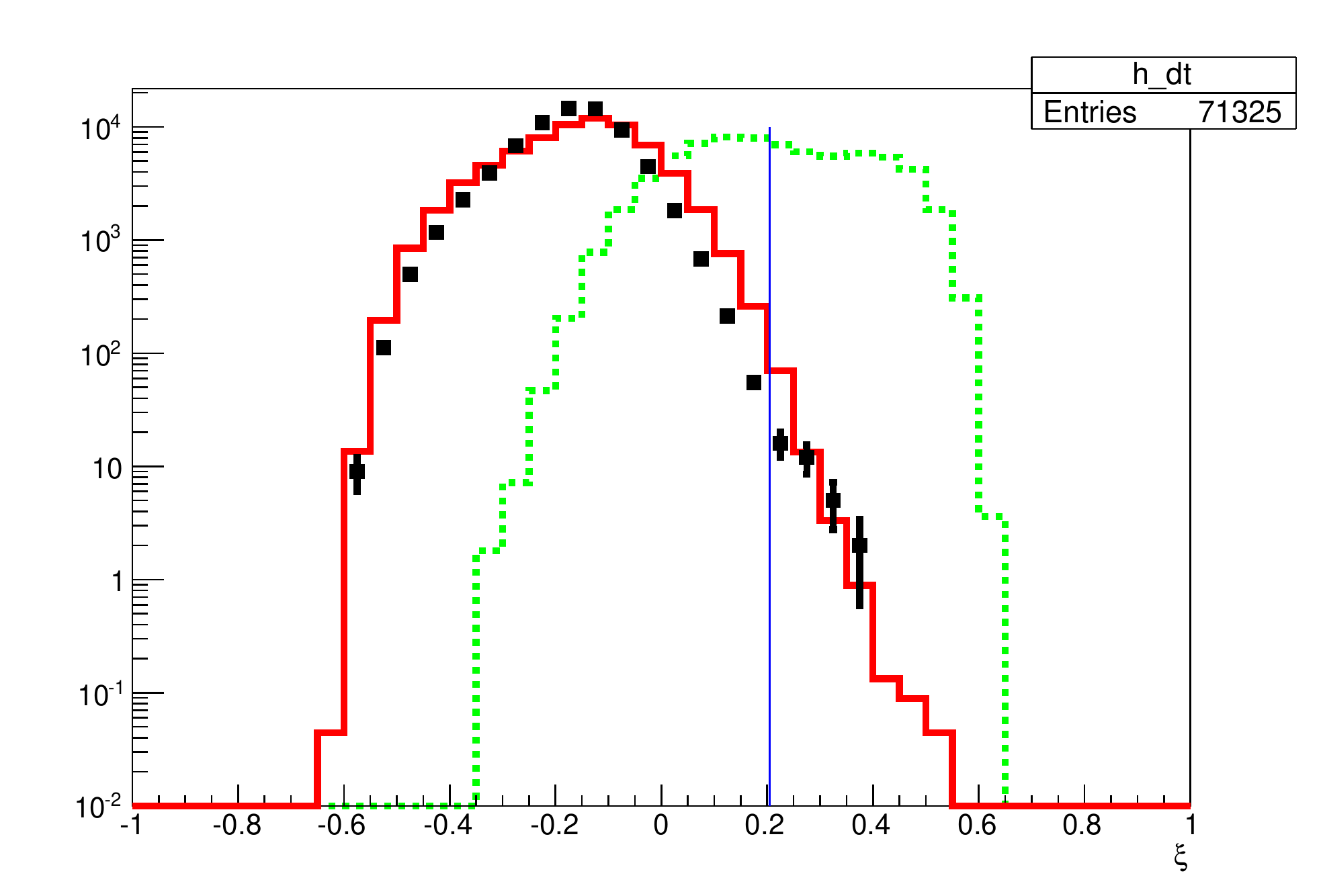}
\caption{The BDT $\xi$ estimator histogram for $E > 10^{18.0}\ \mbox{eV}$. Proton MC is shown with solid red line, neutrino MC is shown with dashed green line, while black dots represent the data. Vertical blue line indicates the median value of $\xi$.}
\label{xihist}
\end{figure}

Figure~\ref{xihist} shows $\xi$ parameter distribution histogram in the energy range $E > 10^{18.0}\ \mbox{eV}$, where proton MC is shown with red line, neutrino MC is shown with green line, while black dots represent the data. One may see that the highly-inclined proton MC set does not follow the data distribution in detail due to the heavier composition as obtained in~\cite{Abbasi:2018wlq} as well as due to the possible systematic effects related to hadronic interaction models, including the “muon excess” problem~\cite{Abbasi:2018fkz}, which is expected to manifest itself more on the higher zenith angles.

\begin{figure}
\includegraphics[width=0.95\columnwidth]{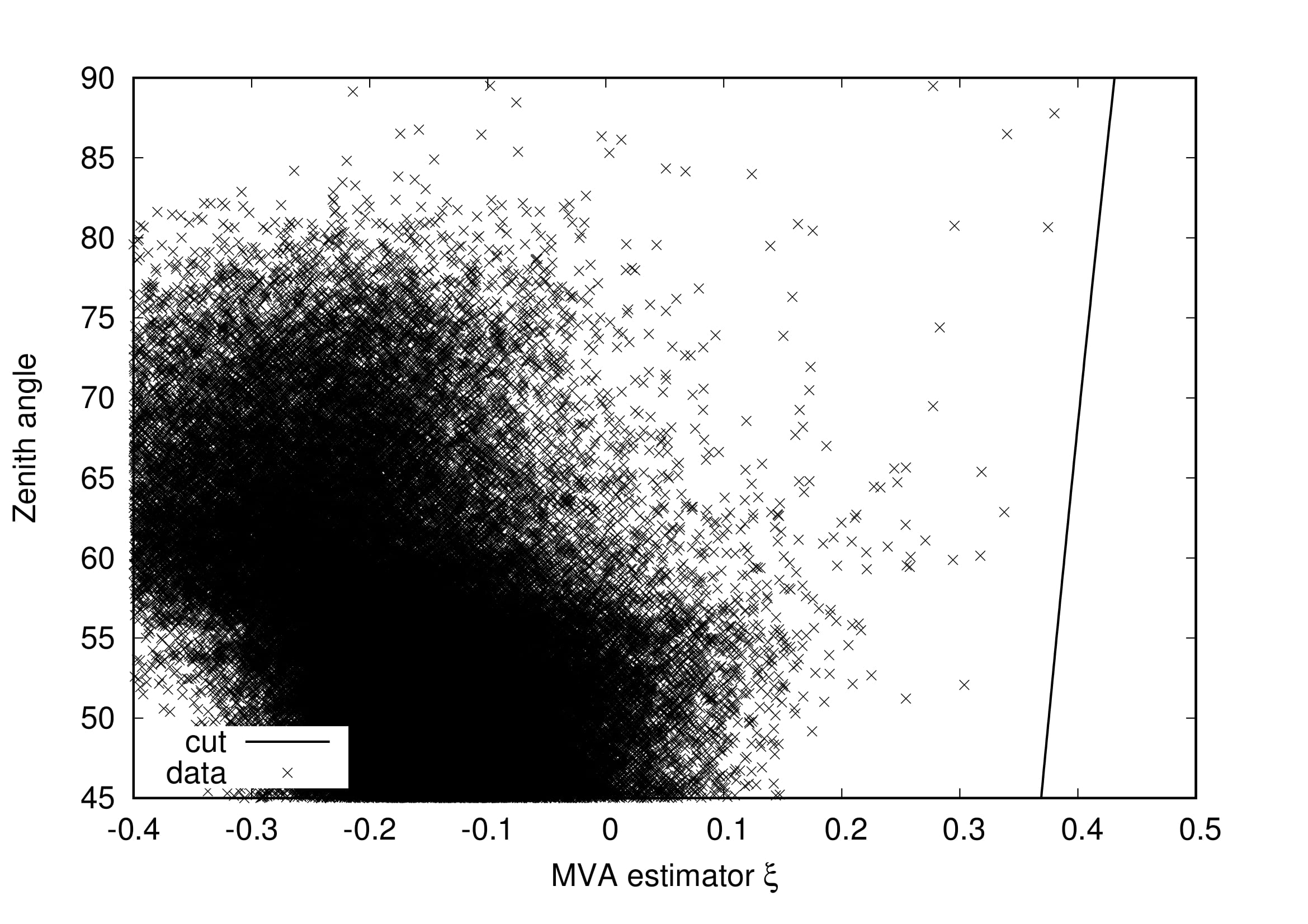}
\caption{The zenith angle distribution for data events plotted against the BDT $\xi$ estimator with the obtained $\xi_{cut}$.}
\label{xiscatter_cut_data}
\end{figure}

The zenith angle-$\xi$ scatter plot for data is shown in Figure~\ref{xiscatter_cut_data} with the obtained $\xi_{cut}$ function.

There are no neutrino candidates in the data set and hence using~\cite{FC}, we derive the upper limit on the number of neutrino events of all flavors in the data: $\overline{n}_{\nu} = 2.44$ at 90\% C.L.

By definition, the integral neutrino flux depends on the number of neutrino events and on the effective neutrino exposure:

\begin{equation}
F_{\nu} = \frac{\overline{n}_{\nu}}{A^{\nu}_{eff}},
\end{equation}

\noindent thus we arrive at the following upper limit on the single flavor diffuse neutrino flux for $E > 10^{18}\ \mbox{eV}$:

\begin{equation*}
E F_{\nu} < 1.85\times10^{-6}\ \mbox{GeV}\ \mbox{cm}^{-2}\ \mbox{s}^{-1}\ \mbox{sr}^{-1}\ \left(90\%\ \mbox{C.L.} \right).
\end{equation*}

\begin{figure}
\includegraphics[width=0.95\columnwidth]{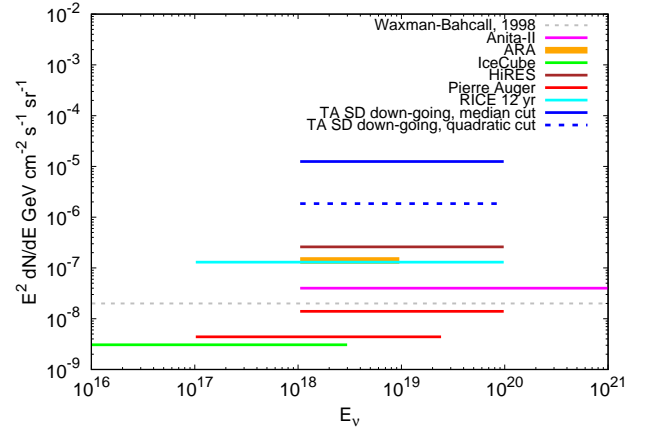}
\caption{The upper limits on the single flavor diffuse differential neutrino flux for $E > 10^{18}\ \mbox{eV}$ obtained with the Telescope Array data in comparison with the Pierre Auger Observatory~\cite{Aab:2019auo}, ANITA-II~\cite{ANITAII}, RICE~\cite{Kravchenko:2011im}, IceCube~\cite{Aartsen:2013dsm}, ARA~\cite{Allison:2019xtn} and HiRes~\cite{Abbasi:2008hr} results along with the Waxman-Bahcall limit~\cite{Waxman:1998yy}.}
\label{nulimit}
\end{figure}

The obtained limit in comparison with the Pierre Auger Observatory~\cite{Aab:2015kma}, ANITA-II~\cite{ANITAII}, RICE~\cite{Kravchenko:2011im}, IceCube~\cite{Aartsen:2013dsm}, ARA~\cite{Allison:2019xtn} and HiRes~\cite{Abbasi:2008hr} results is shown in Figure~\ref{nulimit} as well as the the Waxman-Bahcall limit~\cite{Waxman:1998yy}.

One may be also interested in the results of the neutrino search without the cut optimization procedure explained in~\ref{subsec:FC}. For this purpose, following the line of~\cite{Homola}, we employ the cut based on the median of the $\xi$ distribution for neutrino.

From the distribution of the $\xi$ parameter for neutrino set, see Fig.~\ref{xihist}, one obtains the median value of $\xi_{med} = 0.205$, which corresponds to 50\% of MC neutrino passing the $\xi_{med}$ cut ($N_{pass} = 19856$).

The effective exposure for the down-going neutrino in this case:

\begin{equation*}
\hat{A}^{\nu}_{eff} = 1.2\times 10^{-2}\ {\mbox{km}}^2\ \mbox{sr}\ \mbox{yr}.
\end{equation*}

One obtains $35$ neutrino candidates in the data, which corresponds to the upper limit on neutrino events $\hat{\overline{n}}_{\nu} = 46.4$ at 90\% C.L. It leads to the upper limit on the single flavor diffuse neutrino flux for $E > 10^{18}\ \mbox{eV}$, shown in Fig.~\ref{nulimit}:

\begin{equation*}
E \hat{F}_{\nu} < 4.2\times10^{-6}\ \mbox{GeV}\ \mbox{cm}^{-2}\ \mbox{s}^{-1}\ \mbox{sr}^{-1}\ \left(90\%\ \mbox{C.L.} \right).
\end{equation*}

One should note note that given any fixed cut on $\xi$ one arrives at valid neutrino flux limit, while as expected, the optimization procedure results in the stronger limits. 

Let us finally discuss the effect of the hadronic interaction model systematics and the composition uncertainty on the result of the \textit{Paper}. Both these effects infer the analysis through the Monte-Carlo set used as a background, i.e. the highly-inclined proton MC set. This set is used for the construction of the classifier, including the BDT algorithm and the optimization of the cut. An optimal sensitivity is by construction achieved in the case of a perfect agreement of the data and Monte-Carlo. Incomplete desciption of the observed events by a Monte Carlo set leads either to reduced exposure or to an increased number of false candidates compared to an optimal classifier. Therefore the neutrino flux limit derived is a conseravtive estimate which is close to optimal limit given the reasonable agreement of data and Monte-Carlo demonstrated in the present \textit{Paper}. The sensitivity of the method can be further improved by producing hadronic Monte-Carlo which better reproduces the bulk of data events.

\section*{Acknowledgment}
\label{sec:acknowledgment}

The Telescope Array experiment is supported by the Japan Society for
the Promotion of Science (JSPS) through 
Grants-in-Aid
for Priority Area
431,
for Specially Promoted Research 
JP21000002, 
for Scientific  Research (S) 
JP19104006, 
for Specially Promoted Research 
JP15H05693, 
for Scientific  Research (S)
JP15H05741,
for Science Research (A) JP18H03705 
 and
for Young Scientists (A)
JPH26707011; 
by the joint research program of the Institute for Cosmic Ray Research (ICRR), The University of Tokyo; 
by the U.S. National Science
Foundation awards PHY-0601915,
PHY-1404495, PHY-1404502, and PHY-1607727; 
by the National Research Foundation of Korea
(2016R1A2B4014967, 2016R1A5A1013277, 2017K1A4A3015188, 2017R1A2A1A05071429);
IISN project No. 4.4502.13,
and Belgian Science Policy under IUAP VII/37 (ULB). The development and application of the multivariate
analysis method is supported by the Russian Science Foundation grant
No. 17-72-20291 (INR). The foundations of
Dr. Ezekiel R. and Edna Wattis Dumke, Willard L. Eccles, and George
S. and Dolores Dor\'e Eccles all helped with generous donations. The
State of Utah supported the project through its Economic Development
Board, and the University of Utah through the Office of the Vice
President for Research. The experimental site became available through
the cooperation of the Utah School and Institutional Trust Lands
Administration (SITLA), U.S. Bureau of Land Management (BLM), and the
U.S. Air Force. We appreciate the assistance of the State of Utah and
Fillmore offices of the BLM in crafting the Plan of Development for
the site.  Patrick Shea assisted the collaboration with valuable advice 
on a variety of topics. The people and the officials of Millard County, 
Utah have been a source of
steadfast and warm support for our work which we greatly appreciate. 
We are indebted to the Millard County Road Department for their efforts 
to maintain and clear the roads which get us to our sites. 
We gratefully acknowledge the contribution from the technical staffs of
our home institutions. An allocation of computer time from the Center
for High Performance Computing at the University of Utah is gratefully
acknowledged. The cluster of the Theoretical Division of INR RAS was
used for the numerical part of the work. This work was partially supported by the Collaborative research program of the Institute for Cosmic Ray Research (ICRR), the University of Tokyo.

\end{document}